\documentclass[%
 aip,
 amsmath,amssymb,
 reprint,%
]{revtex4-2}
\usepackage{graphicx}
\usepackage{dcolumn}
\usepackage{bm}

\usepackage[utf8]{inputenc}
\usepackage[T1]{fontenc}
\usepackage{mathptmx}
\usepackage{hyperref}
\hypersetup{colorlinks=true,linkcolor=blue,citecolor=blue, filecolor=blue,urlcolor=blue}
\usepackage{graphicx}

\begin{document}

\preprint{AIP/123-QED}

\title{Gate-controlled analog memcapacitance in LaAlO$_3$/SrTiO$_3$ interface-based devices}

\author{Soumen Pradhan}
\email{soumen.pradhan@uni-wuerzburg.de}
\affiliation{Julius-Maximilians-Universität Würzburg, Physikalisches Institut and Würzburg-Dresden Cluster of Excellence ct.qmat, Am Hubland, 97074 Würzburg, Deutschland}
\author{Victor Lopez-Richard}
\affiliation{Department of Physics, Federal University of São Carlos, 13565-905 São Carlos, SP, Brazil}
\author{Igor Ricardo Filgueira e Silva}
\affiliation{Department of Physics, Federal University of São Carlos, 13565-905 São Carlos, SP, Brazil}
\author{Fabian Hartmann}
\email{fabian.hartmann@uni-wuerzburg.de}
\affiliation{Julius-Maximilians-Universität Würzburg, Physikalisches Institut and Würzburg-Dresden Cluster of Excellence ct.qmat, Am Hubland, 97074 Würzburg, Deutschland}
\author{Ana Luiza Costa Silva}
\affiliation{Department of Physics, Federal University of São Carlos, 13565-905 São Carlos, SP, Brazil}
\author{Leonardo K. Castelano}
\affiliation{Department of Physics, Federal University of São Carlos, 13565-905 São Carlos, SP, Brazil}
\author{Merit Spring}
\affiliation{Julius-Maximilians-Universität Würzburg, Physikalisches Institut and Würzburg-Dresden Cluster of Excellence ct.qmat, Am Hubland, 97074 Würzburg, Deutschland}

\author{Silke Kuhn}
\affiliation{Julius-Maximilians-Universität Würzburg, Physikalisches Institut and Würzburg-Dresden Cluster of Excellence ct.qmat, Am Hubland, 97074 Würzburg, Deutschland}
\author{Michael Sing}
\affiliation{Julius-Maximilians-Universität Würzburg, Physikalisches Institut and Würzburg-Dresden Cluster of Excellence ct.qmat, Am Hubland, 97074 Würzburg, Deutschland}
\author{Ralph Claessen}
\affiliation{Julius-Maximilians-Universität Würzburg, Physikalisches Institut and Würzburg-Dresden Cluster of Excellence ct.qmat, Am Hubland, 97074 Würzburg, Deutschland}
\author{Sven Höfling}
\affiliation{Julius-Maximilians-Universität Würzburg, Physikalisches Institut and Würzburg-Dresden Cluster of Excellence ct.qmat, Am Hubland, 97074 Würzburg, Deutschland}

\date{\today}

\begin{abstract}
Current memcapacitor implementations typically demand complex fabrication processes or depend on organic materials exhibiting poor environmental stability and reproducibility. Here, we demonstrate memcapacitor structures utilizing a quasi 2-dimensional electron gas, formed at the crystalline LaAlO$_3$/SrTiO$_3$ heterointerface, as electrodes and SiO$_2$/SrTiO$_3$ as dielectric layer. The observed memcapacitance originates from the charge localization in a lateral floating gate, while an applied gate voltage enables reversible tuning of the device capacitance. Furthermore, preprogrammed or erased gate biases enable controllable shifts of the capacitance hysteresis window toward positive or negative bias, leading to an enlarged capacitance gap at zero bias. A memcapacitor model developed for this system reproduces the main features of the experimental capacitance hysteresis, capturing the effects of charge fluctuations and dielectric frequency modulation within the oxide layer. The demonstrated low-voltage operation and gate tunability of oxide interface-based memcapacitors highlight their potential for power-efficient, capacitor-based neuromorphic and synaptic electronic architectures.
\end{abstract}

\maketitle
Human brain-inspired neuromorphic computing technologies play a crucial role in addressing the increasing demands of advanced information processing systems. The field gained significant attention following the experimental realization of memristive devices~\cite{strukov2008missing}. More recently, memcapacitors--whose capacitance depends on the history of the applied voltage or charge--have attracted increasing interest as alternative mem-elements for neuromorphic computing due to their ultralow static power consumption compared to memristors~\cite{demasius2021energy,pei2023power,bhardwaj2025toward,kim2025voltage}. Furthermore, memcapacitors can be integrated as the gate stack in metal-oxide-semiconductor field-effect transistors (MOSFETs), where capacitance modulation effectively controls the drain current~\cite{yang2016memcapacitive,park2018analog}. Consequently, memcapacitor-based MOSFETs present a promising approach to replace conventional flash memory technologies, in which a charge storage node shifts the subthreshold voltage to enable memory operation.

To date, only a limited number of theoretical and experimental studies have reported on memcapacitor devices. The approaches explored so far include variations in effective dielectric thickness\cite{emara2017non} and area\cite{wang2018capacitive}, dielectric permittivity\cite{8733844} and ion migration within the dielectric layer\cite{lai2010ionic}. Among these, charge-trap-based memory devices stand out due to their easy scalability, compatibility with large-scale fabrication, and potential for high-density, low-cost data storage\cite{hwang2023memcapacitor,8970565}. For instance, Hwu \textit{et al.} reported capacitance hysteresis in SiO$_2$-based concentric metal-oxide-semiconductor (MOS) capacitors induced by charge trapping in ultrathin SiO$_2$\cite{chen2014effect}, charge modulation by outer MOS-gate\cite{8608000} and, more recently capacitive synapses using lateral coupling with high sensitivity in the inversion regime\cite{kao2025inversion}. Beyond realizations based on inorganic materials, memcapacitors have also been engineered from organic thin films utilizing charge trapping in polymer electret layers\cite{cai2019organic}. Interestingly partial coverage of the electret layer enables multiple capacitive states providing analog and nonvolatile memory functionality\cite{9845432}.

As a new player in the game, LaAlO$_3$/SrTiO$_3$ (LAO/STO) heterostructures with quasi 2-dimensional electron gas (q2-DEG) formed at the interface\cite{ohtomo2004high} exhibit exceptionally high gate capacitance underscoring their potential for low power electronic applications\cite{li2011very}. To date, only a few studies have reported capacitance memory effects in LAO/STO systems, typically attributed to structural distortions\cite{bi2016electro}, migration of oxygen vacancies\cite{wu2013electrically} or deep trap states\cite{kim2015electric} and yet these mechanisms remain insufficiently explored. In this letter, we have designed and fabricated lateral nanoelectronic devices based on LAO/STO heterostructures to systematically investigate memcapacitance behavior. By employing an auxiliary control gate, operated either at floating potential or under an applied bias, we demonstrate tunable capacitance memory and reversible non-linear capacitance characteristics thereby advancing the prospects of capacitance-based artificial neural networks.

\begin{figure}
\includegraphics[width=0.48\textwidth]{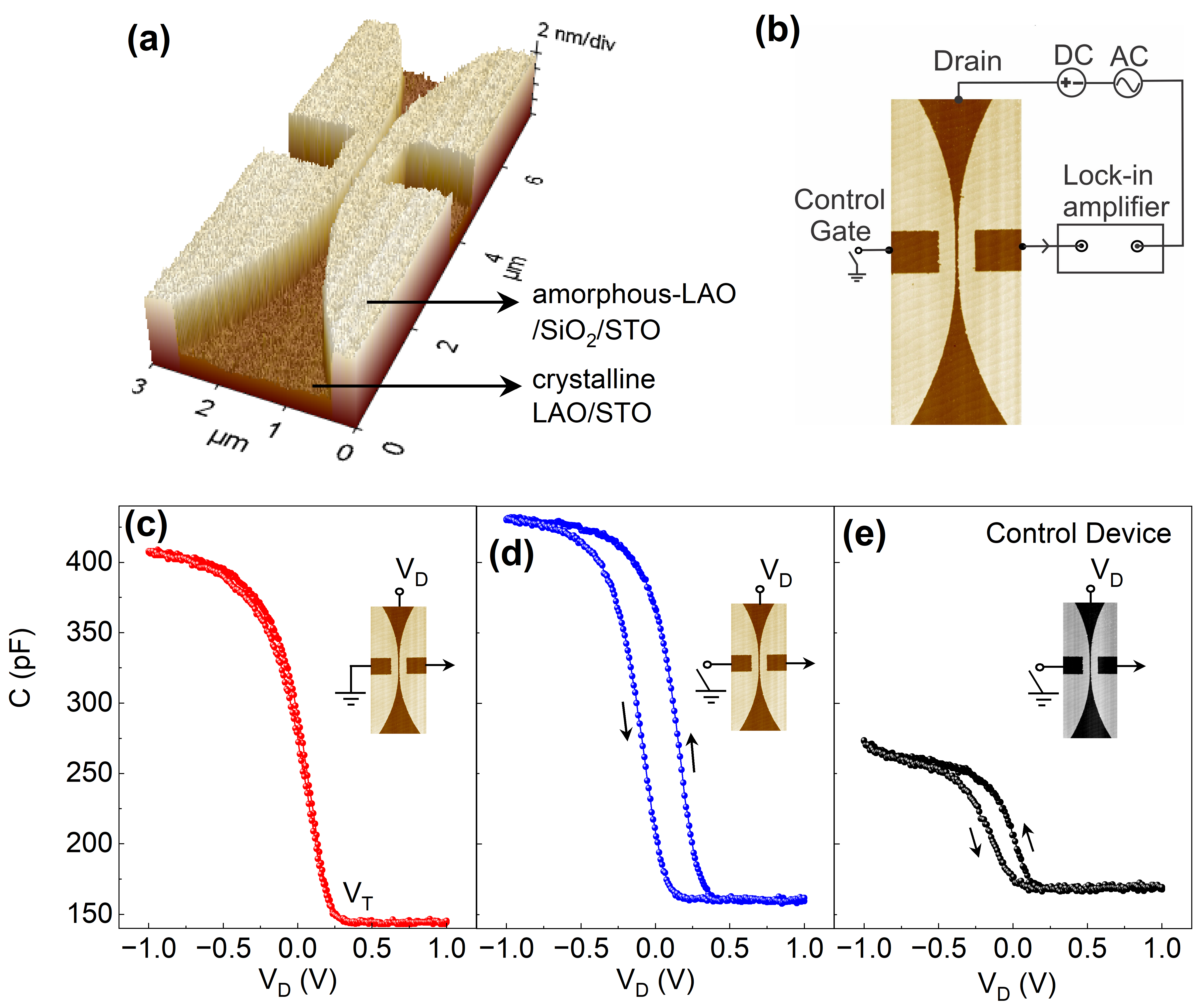}
\caption{\label{fig:f1} Capacitance memory in LAO/STO interface-based lateral nanoelectronic device: (a) 3-dimensional atomic force microscopy (AFM) image of the nanowire device with two lateral gates. (b) 2-dimensional AFM image of the device with circuit diagram for capacitance measurement. C-V hysteresis curves with drain voltage (V$_D$) sweep between $\pm$1 V at an AC voltage of frequency 10 Hz with control gate (c) grounded and (d) at floating condition for original device and (e) at floating condition for a control device of same dimensions without SiO$_2$ and amorphous LAO layer on top of dielectric layer. The insets show the AFM images of the devices with gating configurations for the corresponding measurements.}
\end{figure}

To fabricate the device, a TiO$_2$ terminated STO single crystalline substrate was first patterned with a negative photoresist using electron beam lithography. Then, a 11 nm-thick SiO$_2$ layer was grown using thermal evaporation followed by a lift-off process to create well defined insulating areas. Finally, a 6 unit-cell-thick LAO film was grown over the entire surface using pulsed laser deposition (PLD). The process resulted in crystalline LAO growth directly on the exposed STO regions and amorphous LAO growth on areas covered with SiO$_2$, as shown as topographically low and high regions, respectively, in the 3-dimensional atomic force microscopy (AFM) image displayed in Fig. \hyperref[fig:f1]{1(a)}. Therefore, q2-DEG forms at the crystalline LAO/STO interface, while the amorphous LAO/SiO$_2$ regions remain insulating. The resulting lateral capacitor structure comprises the q2-DEG nanowire channel and a lateral gate ($\approx$1 $\mu$m wide) as conducting lateral electrodes separated by a $\approx$350 nm-wide SiO$_2$ dielectric layer. An identical lateral gate on the opposite side, also based on the q2-DEG, serves as a control gate with SiO$_2$/SrTiO$_3$ acting as the gate dielectric. Further details on the fabrication procedure and growth parameters can be found in our recent report\cite{pradhan2025oxide}. For the capacitance measurement, an AC voltage of 20 mV amplitude superimposed on a DC bias was applied to the drain contact using a function generator (Model: Keithley 3390). The resulting current output from one lateral gate was fed into a lock-in amplifier (EG\&G Instruments, model: 7265) while the same input signal was used as reference. The real and imaginary components of the current were recorded to extract the device capacitance. The experimental setup and corresponding two-dimensional AFM image of the device are shown in Fig. \hyperref[fig:f1]{1(b)}. 

Figure~\hyperref[fig:f1]{1(c)} shows the capacitance–voltage (C–V) characteristics measured between the drain and one lateral gate at an AC frequency of 10 Hz, with the drain voltage ($V_D$) swept between $\pm$1 V at a rate of 26.66 mV/s while keeping the control gate grounded. A step-like behavior in capacitance is observed, characterized by a high-capacitance state ($C_{\text{high}}$) at negative voltages and a low-capacitance state ($C_{\text{low}}$) at positive voltages, without hysteresis. The bias voltage at which the capacitance begins to increase from $C_{\text{low}}$ is defined as the threshold voltage ($V_T$). Here, $C_{\text{high}}$ corresponds to the accumulation regime, dominated by the oxide layer, whereas $C_{\text{low}}$ arises in the depletion regime, where the series combination of depletion and oxide capacitances reduces the total capacitance. When the control gate is left floating, the device exhibits pronounced C–V hysteresis loops, as shown in Fig.~\hyperref[fig:f1]{1(d)}. The hysteresis originates from charge localization on the floating control gate, which induces a shift in $V_T$ between forward and reverse sweeps, i.e., the signature of memcapacitance. To further confirm the origin of memcapacitance, a control device of identical geometry was fabricated by etching the LAO layer from predefined regions of a crystalline LAO/STO sample with the same LAO thickness (6 unit cells), thereby avoiding the SiO$_2$ and amorphous LAO regions present in the original device. The C–V characteristics of the control device were measured using the same protocol, as shown in Fig.~\hyperref[fig:f1]{1(e)}. Insets show 2-dimensional AFM images of the devices with the control gate configurations used for the measurements. The control sample exhibits similar $C_{\text{low}}$ values but a lower $C_{\text{high}}$, attributable to the reduced effective dielectric constant due to the presence of air above the STO surface in the accumulation state. The persistence of hysteresis in the control sample confirms that the memcapacitance originates from charge localization on the control gate, ruling out contributions from traps in the amorphous LAO or SiO$_2$.

\begin{figure}
\includegraphics[width=0.48\textwidth]{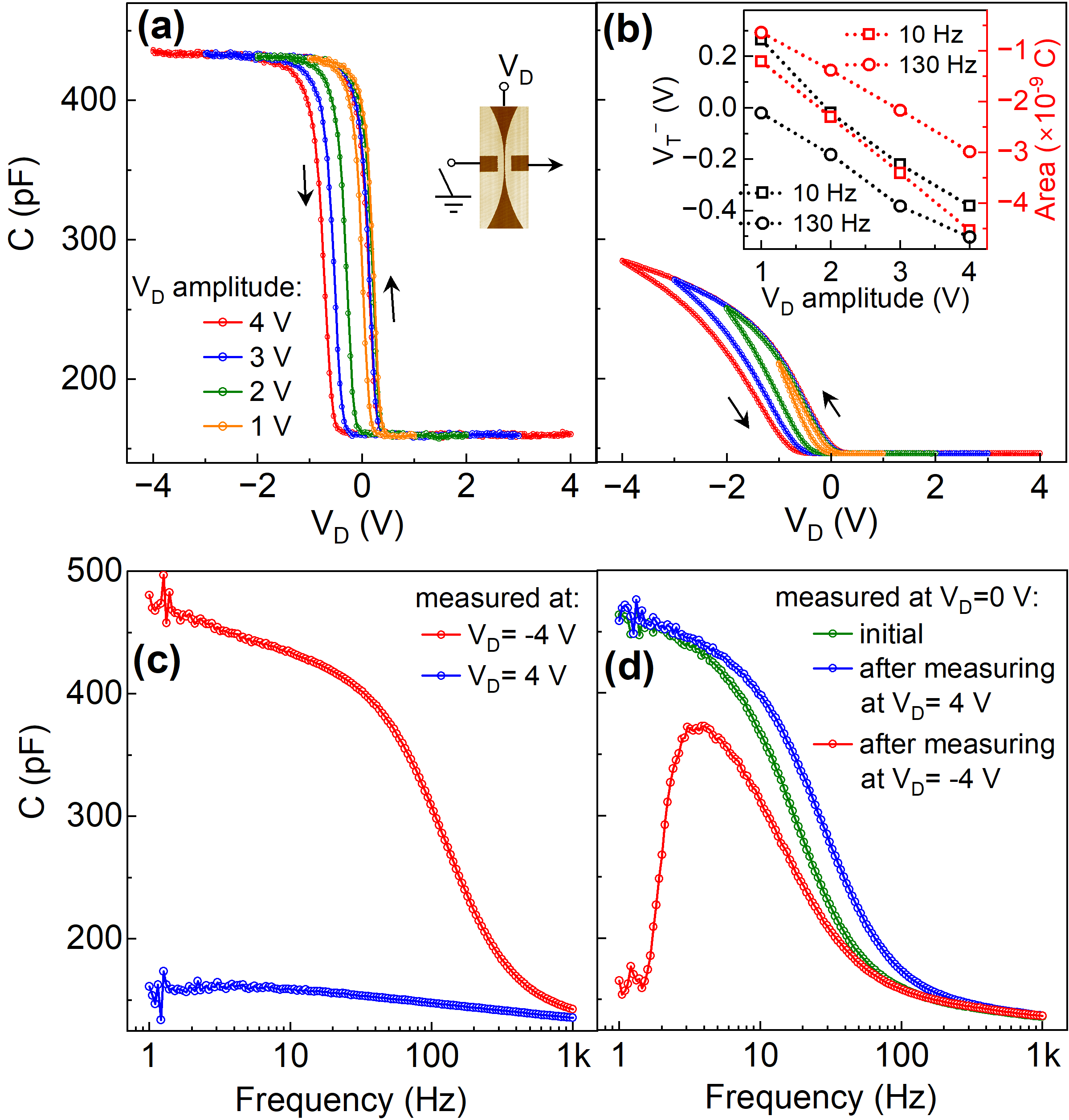}
\caption{\label{fig:f2} Amplitude and frequency modulation of capacitance with floating control gate: C-V hysteresis curves for different drain voltage (V$_D$) amplitude with control gate at floating condition measured at (a) 10 Hz and (b) 130 Hz with inset showing variation of hysteresis area (right) and negative threshold voltage (V$_T^-$) (left) with V$_D$ amplitude extracted from (a) and (b). (c) capacitance-frequency (C-f) plots at V$_D$ of 4 and -4 V, (d) C-f plots measured at V$_D$ of 0 V including its initial state, after measuring at V$_{D}$=4 V and -4 V  showing memory in capacitance in the frequency spectrum.}
\end{figure}
To gain deeper insight into the characteristics of the original device, amplitude- and frequency-dependent capacitance modulation were investigated. Figures~\hyperref[fig:f2]{2(a)} and~\hyperref[fig:f2]{2(b)} show the C–V hysteresis loops for the device at 10 Hz and 130 Hz, respectively, using a voltage sweep sequence of $+V \rightarrow -V \rightarrow +V$ for various amplitudes with the control gate left floating. At 10 Hz, both $C_{\text{high}}$ and $C_{\text{low}}$ remain unchanged even when $V_D$ is increased to 4 V. In contrast, at 130 Hz, $C_{\text{high}}$ increases gradually with increasing $V_D$, while $C_{\text{low}}$ remains constant. This behavior reflects frequency-dependent dipole dynamics: at low frequencies, dipoles can fully align even under small biases, whereas at high frequencies their response lags behind the applied field, resulting in a bias-dependent increase in the effective dipole moment. The bias at which the capacitance sharply rises during the reverse sweep ($V_T^+$) remains nearly constant for all voltage amplitudes and frequencies. In contrast, during the forward sweep, the threshold voltage ($V_T^-$) shifts systematically toward more negative values with increasing voltage amplitude. The variation of $V_T^-$ and the corresponding hysteresis area with voltage amplitude are shown in the inset of Fig.~\hyperref[fig:f2]{2(b)} for both frequencies, showing an almost linear dependence of $V_T^-$ on the applied amplitude and a strongly linear increase of the hysteresis area. For further analysis, capacitance–frequency (C–f) curves were measured at $V_D = 4$ V and $-4$ V with the control gate left floating, as shown in Fig.~\hyperref[fig:f2]{2(c)}. At $V_D = 4$ V, the capacitance remains nearly constant at $\sim$150 pF across the entire frequency range. In contrast, at $V_D = -4$ V, a pronounced enhancement in capacitance is observed at low frequencies, consistent with Figs.~\hyperref[fig:f2]{2(a)} and~\hyperref[fig:f2]{2(b)}. This frequency-dependent modulation at negative bias can be partially attributed to the tuning of the effective dielectric function: as frequency increases, the capacitance progressively decreases, approaching $C_{\text{low}}$ near 1 kHz through approximately two reduction rates. To further probe the memory behavior, C–f measurements were performed at $V_D = 0$ V in three conditions: initial state, after applying $V_D = 4$ V, and after applying $V_D = -4$ V, as shown in Fig.~\hyperref[fig:f2]{2(d)}. The results reveal a high capacitance at zero bias across all frequencies following a 4 V measurement and a low capacitance after a -4 V measurement, consistent with the direction of the hysteresis loop. In contrast, the initial state exhibits an intermediate capacitance over the full frequency range. These observations confirm that the device exhibits robust capacitance memory at low frequencies, which gradually diminishes as frequency increases.

\begin{figure}
\includegraphics[width=0.48\textwidth]{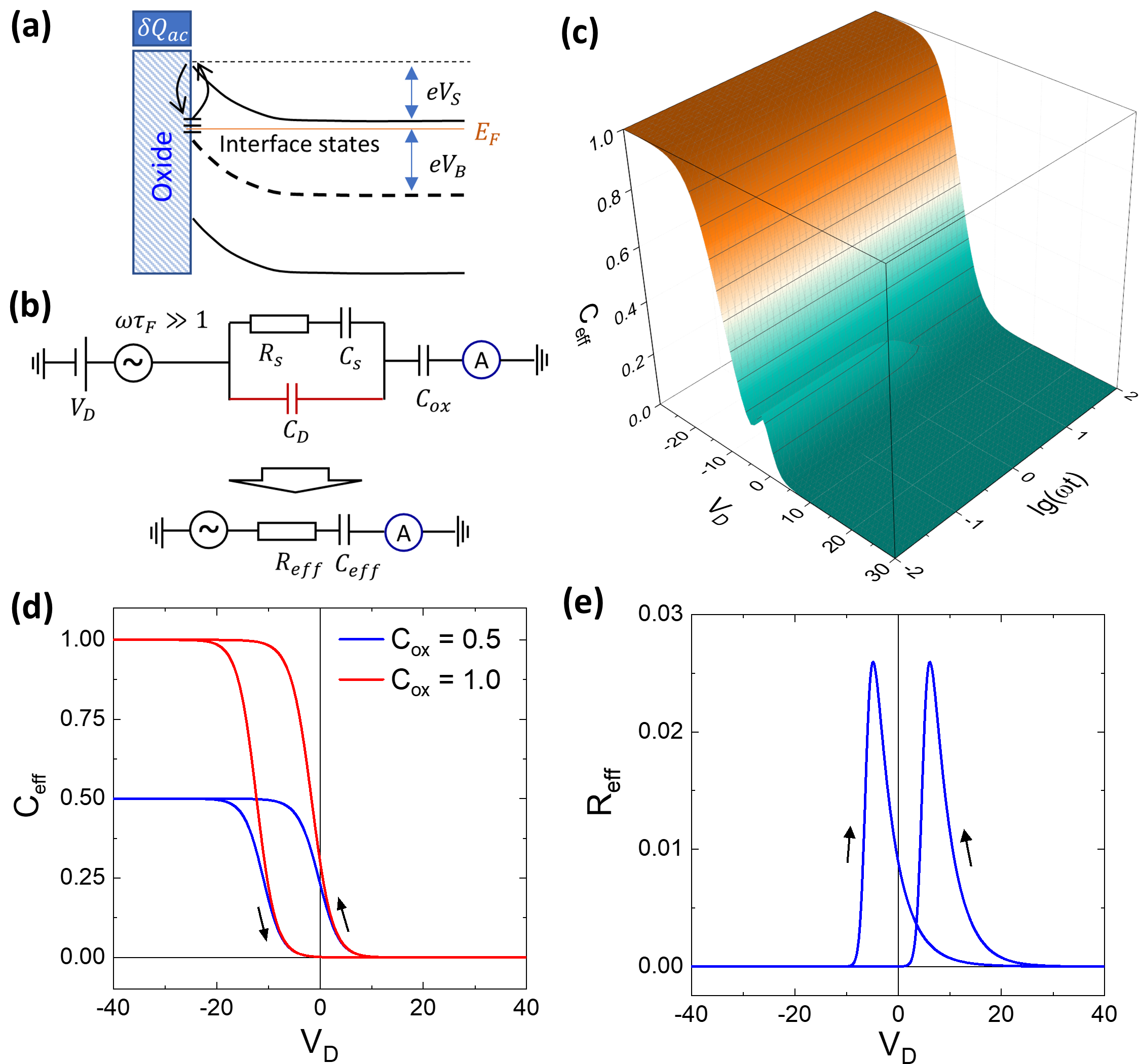}
\caption{\label{fig:f3} (a) Schematic band diagram of the oxide layer and the adjacent bent bands in the depleted semiconductor region of the device, with interface states involved in electronic transitions indicated by arrows. (b) Equivalent circuit representing the carrier dynamics, and its reduction to a simpler series combination of an effective resistance and an effective capacitance. (c) Calculated effective capacitance as a function of the applied DC bias and the driving frequency of the AC component. (d) Simulated effective capacitance as a function of bias potential for two different values of the oxide-layer capacitance and (e) corresponding effective resistance calculated under the same conditions as in (d)}
\end{figure}
To elucidate the device characteristics, first we employ a theoretical capacitor model adapted from the classical framework introduced in Ref.~\citenum{Nicollian1967}. In this approach, the effective capacitance is governed by charge fluctuations activated within the dielectric (oxide) layer of the capacitor, which couple to the interfacial states and modulate the overall carrier dynamics, as represented in Fig. \hyperref[fig:f3]{3(a)}. Under a voltage difference between the drain and the gate, consisting of a DC bias $V_D$ modulated by an AC component $\delta V_t$, the apparent impedance can be represented by the equivalent circuit shown in the top panel of Fig. \hyperref[fig:f3]{3(b)} and expressed as
\begin{equation}
    Z_T = \frac{1}{Y_s + i \omega C_D} + \frac{1}{i \omega C_{ox}},
\end{equation}
where $C_{ox}$ is the capacitance of the oxide layer, $C_D$ is the depletion-region capacitance, and $Y_s$ is the admittance associated with carrier capture–emission dynamics at the oxide/semiconductor interface~\cite{Nicollian1967}, given by
\begin{equation}
    Y_s = \frac{1}{R_s - \frac{i}{\omega C_s}}.
\end{equation}
Here, $C_s = \frac{e^2 N_s f_0(V_s)[1-f_0(V_s)]}{k_B T}$ and $R_s = \tau / C_s$, with $N_s$ denoting the density of interface trap states (defects) and $f_0(V_s)$ the occupation probability determined by the surface potential
\[
    V_s = \eta V_D + \frac{\delta Q_{ac}}{C_{ac} A}.
\]
In this expression, $\delta Q_{ac}$ denotes the slowly varying charge fluctuations in the floating gate (relative to the AC frequency $\omega$), while $\eta$ represents the fraction of the total voltage drop appearing at the surface potential. For the sake of a qualitative analysis, $\eta$ is treated here as a constant. The total apparent impedance can thus be reduced to a simpler equivalent circuit consisting of an effective resistance in series with an effective capacitance, as represented in the lower panel of Fig. \hyperref[fig:f3]{3(b)}. These quantities are defined as 
\[
    R_{\mathrm{eff}} = \Re \big( Z_T \big), 
    \qquad 
    C_{\mathrm{eff}} = -\frac{1}{\omega \, \Im \big( Z_T \big)},
\]
which allows direct correlation of the model with the effective parameters obtained experimentally. The resulting effective capacitance, normalized to the oxide-layer capacitance, is shown in Fig. \hyperref[fig:f3]{3(c)} as a function of the AC frequency and the applied DC bias. The oxide-layer capacitance defines the upper bound for the total effective capacitance, which decreases with increasing bias due to the expansion of the depletion region in the semiconductor. At low frequencies ($\omega \tau < 1$), an additional capacitive contribution arises from the delayed response of trapped carriers at the interface.

To analyze the origin of the capacitance hysteresis we consider that the fluctuating charge in the floating gate evolves according to the relaxation-time approximation described in Ref.\citenum{LopezRichard2022},
\[
    \frac{d \, \delta Q_{ac}}{dt} = -\frac{\delta Q_{ac}}{\tau} + g(V_D),
\]
where $g(V_D)$ represents a suitable generation function. Within this framework, the effective capacitance develops a hysteresis, as shown in Fig.~\hyperref[fig:f3]{3(d)}, in close correspondence with the experimental observations. To assess the influence of the oxide dielectric properties, two different values of $C_{ox}$ were considered, thereby examining how variations in dielectric capacitance contribute to the overall device response. The interfacial charge dynamics affect not only the capacitance but also induce a bias-dependent modulation of the resistance, which cannot be neglected. This effect has a direct impact on the overall device performance, as illustrated in Fig.~\hyperref[fig:f3]{3(e)}. To verify the presence of this resistance hysteresis we also measured the resistance using the same voltage sweep protocol employed for the capacitance measurement.The resulting hysteresis loops, consistent with the model predictions, are presented in supplementary Fig.~S1.

\begin{figure}
\includegraphics[width=0.48\textwidth]{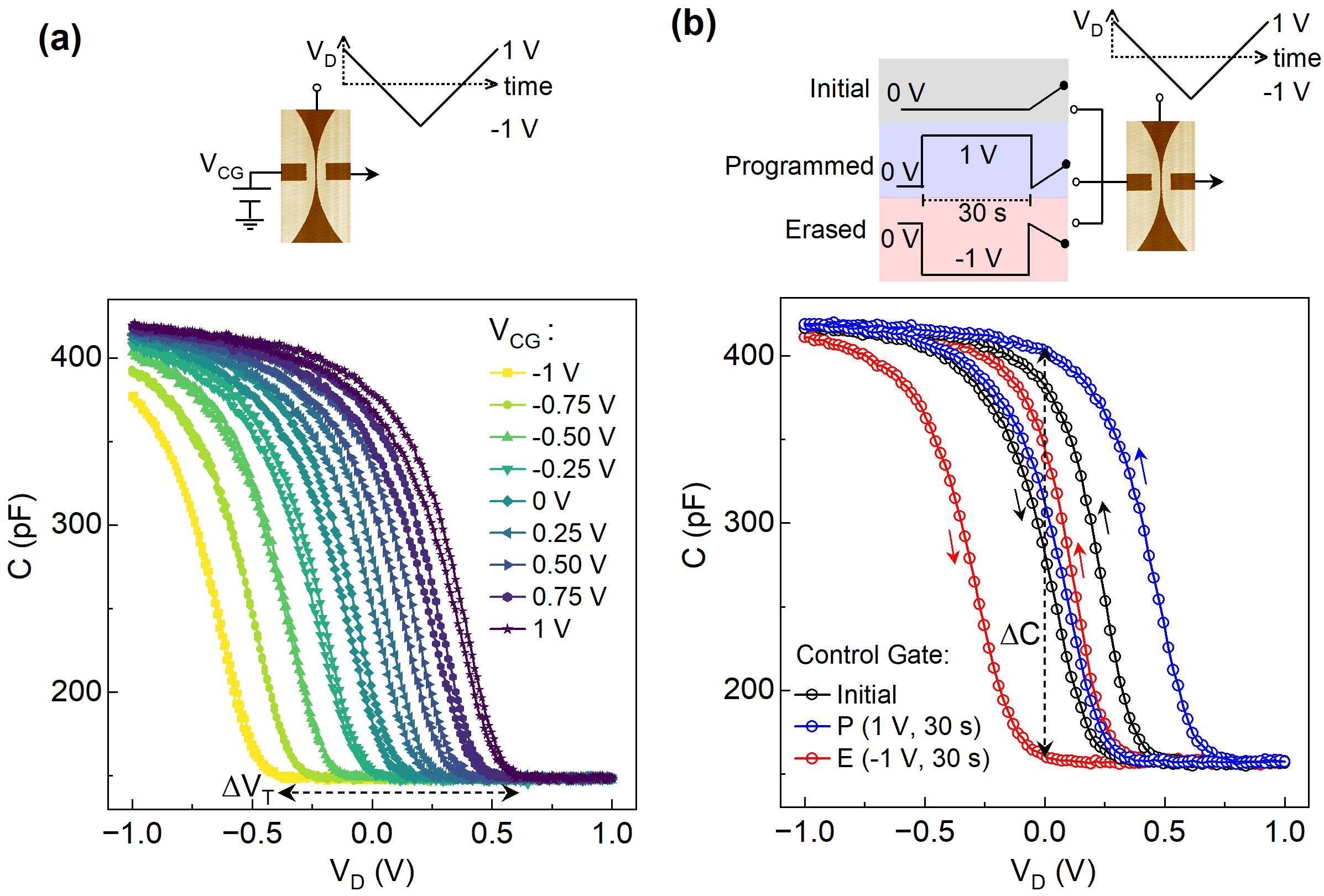}
\caption{\label{fig:f4} Tuning of reversible capacitance and its memory window using control gate: (a) C-V reversible curves with full cycle V$_D$ sweep between $\pm$1 V for different control gate voltage (V$_{CG}$) showing positive and negative shift of threshold voltage (V$_T$), (b) C-V hysteresis curves for same V$_D$ range with control gate at floating condition including its initial state, programmed state (P) after programming with V$_{CG}$=1 V for 30 s and erased state (E) after erasing with V$_{CG}$=-1 V for 30 s, showing flexibility to tune the capacitance gap at V$_{D}$=0 V. The schematic representation for the measurements are shown on top panel.}
\end{figure}

To investigate the tunability of capacitance using the control gate, full C–V sweep measurements between $\pm$1 V were performed for various control gate voltages ($V_{CG}$) ranging from -1 to 1 V, as shown in Fig.~\hyperref[fig:f4]{4(a)}. The $V_T$ shifts systematically, with a maximum tunability ($\Delta V_T$) of $\sim$1 V, without the emergence of memory~\cite{cai2019organic}. The extracted $V_T$ values and their dependence on $V_{CG}$ are presented in Supplementary Fig.~S2. At negative $V_{CG}$, additional electron accumulation at the control gate restricts charge flow through the drain, requiring a more negative $V_D$ to achieve accumulation. Conversely, positive $V_{CG}$ facilitates accumulation, yielding positive $V_T$ values that gradually saturate at higher $V_{CG}$. To further understand the nonlinear C–V characteristics, charge–voltage (Q–V) curves were obtained by integrating the C–V data, as shown in Supplementary Fig.~S3. The state-dependent charge distribution highlights the potential of MOS capacitor-based crossbar arrays with NAND flash-like architecture, where the synaptic weight of each cell corresponds to the integrated charge ($\int C\,dV$) used in vector–matrix multiplication, as reported by Hwang \textit{et al.}~\cite{hwang2023memcapacitor}. Additional C–V measurements performed by varying $V_{CG}$ at $V_D = 0$ V (Supplementary Fig.~S4) reveal a continuous modulation of capacitance from $\sim$147 pF to $\sim$386 pF within a narrow $V_{CG}$ range of $\pm$1 V, demonstrating the existence of multiple capacitance states. Finally, to explore the effect of the control gate on memcapacitance, C–V hysteresis loops were measured under three conditions: (i) the initial state with a floating control gate, (ii) a programmed state after applying +1 V for 30 s, and (iii) an erased state after applying –1 V for 30 s, as shown schematically in the top panel of Fig.~\hyperref[fig:f4]{4(b)}. The hysteresis loop shifts toward positive bias in the programmed state and toward negative bias in the erased state, both exhibiting larger loop areas compared to the initial condition. Moreover, at zero bias, the capacitance gap ($\Delta C$) increases from approximately 100 pF in the initial state to a maximum of about 243 pF between the programmed and erased states. These results demonstrate that control-gate-induced tuning of the memcapacitance provides an additional degree of freedom for achieving linearity and precision in capacitor-based synaptic weight modulation, an essential feature for neuromorphic computing architectures.

In summary, we demonstrate charge-localization-mediated analog memcapacitance in lateral LAO/STO-based nanoelectronic devices, where gate control enables reversible and tunable capacitance states. The observed hysteresis, frequency dependence, and bias-driven modulation are consistently captured by a model incorporating interfacial charge dynamics and dielectric fluctuations, establishing a clear correspondence between experiment and theory. These findings highlight the potential of oxide memcapacitors as energy-efficient building blocks for synaptic electronics, where the coexistence of volatile and tunable capacitive responses provides enhanced flexibility for neuromorphic and adaptive computing architectures.

The additional supporting results are provided in the Supplementary Material.
\begin{acknowledgments}
The authors gratefully acknowledge financial support from the state of Bavaria and the Deutsche Forschungsgemeinschaft (DFG, German Research Foundation) under Germany’s Excellence Strategy through the Würzburg-Dresden Cluster of Excellence on Complexity and Topology in Quantum Matter “ct.qmat” (EXC 2147, Project ID 390858490) as well as through the Collaborative Research Center SFB 1170 “ToCoTronics” (Project ID 258499086). V.L.R. acknowledges the support from Conselho Nacional de Desenvolvimento Científico e Tecnológico (CNPq-Brasil) Proj. 311536/2022-0. V.L.R and L.K.C. acknowledge FAPESP Projs. 2024/09298-7, 2025/00677-8, and 2025/04805-0. V.L.R and A.L.C.S. acknowledge FAPESP Projs. 2023/17490-2 and 2024/17787-8.
\end{acknowledgments}

\bibliographystyle{aipnum4-1}
\bibliography{aipsamp}

\begin{thebibliography}{26}%
\makeatletter
\providecommand \@ifxundefined [1]{%
 \@ifx{#1\undefined}
}%
\providecommand \@ifnum [1]{%
 \ifnum #1\expandafter \@firstoftwo
 \else \expandafter \@secondoftwo
 \fi
}%
\providecommand \@ifx [1]{%
 \ifx #1\expandafter \@firstoftwo
 \else \expandafter \@secondoftwo
 \fi
}%
\providecommand \natexlab [1]{#1}%
\providecommand \enquote  [1]{``#1''}%
\providecommand \bibnamefont  [1]{#1}%
\providecommand \bibfnamefont [1]{#1}%
\providecommand \citenamefont [1]{#1}%
\providecommand \href@noop [0]{\@secondoftwo}%
\providecommand \href [0]{\begingroup \@sanitize@url \@href}%
\providecommand \@href[1]{\@@startlink{#1}\@@href}%
\providecommand \@@href[1]{\endgroup#1\@@endlink}%
\providecommand \@sanitize@url [0]{\catcode `\\12\catcode `\$12\catcode
  `\&12\catcode `\#12\catcode `\^12\catcode `\_12\catcode `\%12\relax}%
\providecommand \@@startlink[1]{}%
\providecommand \@@endlink[0]{}%
\providecommand \url  [0]{\begingroup\@sanitize@url \@url }%
\providecommand \@url [1]{\endgroup\@href {#1}{\urlprefix }}%
\providecommand \urlprefix  [0]{URL }%
\providecommand \Eprint [0]{\href }%
\providecommand \doibase [0]{http://dx.doi.org/}%
\providecommand \selectlanguage [0]{\@gobble}%
\providecommand \bibinfo  [0]{\@secondoftwo}%
\providecommand \bibfield  [0]{\@secondoftwo}%
\providecommand \translation [1]{[#1]}%
\providecommand \BibitemOpen [0]{}%
\providecommand \bibitemStop [0]{}%
\providecommand \bibitemNoStop [0]{.\EOS\space}%
\providecommand \EOS [0]{\spacefactor3000\relax}%
\providecommand \BibitemShut  [1]{\csname bibitem#1\endcsname}%
\let\auto@bib@innerbib\@empty
\bibitem [{\citenamefont {Strukov}\ \emph {et~al.}(2008)\citenamefont
  {Strukov}, \citenamefont {Snider}, \citenamefont {Stewart},\ and\
  \citenamefont {Williams}}]{strukov2008missing}%
  \BibitemOpen
  \bibfield  {author} {\bibinfo {author} {\bibfnamefont {D.~B.}\ \bibnamefont
  {Strukov}}, \bibinfo {author} {\bibfnamefont {G.~S.}\ \bibnamefont {Snider}},
  \bibinfo {author} {\bibfnamefont {D.~R.}\ \bibnamefont {Stewart}}, \ and\
  \bibinfo {author} {\bibfnamefont {R.~S.}\ \bibnamefont {Williams}},\ }\href
  {\doibase https://doi.org/10.1038/nature06932} {\bibfield  {journal}
  {\bibinfo  {journal} {Nature}\ }\textbf {\bibinfo {volume} {453}},\ \bibinfo
  {pages} {80} (\bibinfo {year} {2008})}\BibitemShut {NoStop}%
\bibitem [{\citenamefont {Demasius}, \citenamefont {Kirschen},\ and\
  \citenamefont {Parkin}(2021)}]{demasius2021energy}%
  \BibitemOpen
  \bibfield  {author} {\bibinfo {author} {\bibfnamefont {K.-U.}\ \bibnamefont
  {Demasius}}, \bibinfo {author} {\bibfnamefont {A.}~\bibnamefont {Kirschen}},
  \ and\ \bibinfo {author} {\bibfnamefont {S.}~\bibnamefont {Parkin}},\ }\href
  {https://www.nature.com/articles/s41928-021-00649-y} {\bibfield  {journal}
  {\bibinfo  {journal} {Nat. Electron.}\ }\textbf {\bibinfo {volume} {4}},\
  \bibinfo {pages} {748} (\bibinfo {year} {2021})}\BibitemShut {NoStop}%
\bibitem [{\citenamefont {Pei}\ \emph {et~al.}(2023)\citenamefont {Pei},
  \citenamefont {Zhu}, \citenamefont {Liu}, \citenamefont {Cui}, \citenamefont
  {Li}, \citenamefont {Yan}, \citenamefont {Li}, \citenamefont {Wan},\ and\
  \citenamefont {Wan}}]{pei2023power}%
  \BibitemOpen
  \bibfield  {author} {\bibinfo {author} {\bibfnamefont {M.}~\bibnamefont
  {Pei}}, \bibinfo {author} {\bibfnamefont {Y.}~\bibnamefont {Zhu}}, \bibinfo
  {author} {\bibfnamefont {S.}~\bibnamefont {Liu}}, \bibinfo {author}
  {\bibfnamefont {H.}~\bibnamefont {Cui}}, \bibinfo {author} {\bibfnamefont
  {Y.}~\bibnamefont {Li}}, \bibinfo {author} {\bibfnamefont {Y.}~\bibnamefont
  {Yan}}, \bibinfo {author} {\bibfnamefont {Y.}~\bibnamefont {Li}}, \bibinfo
  {author} {\bibfnamefont {C.}~\bibnamefont {Wan}}, \ and\ \bibinfo {author}
  {\bibfnamefont {Q.}~\bibnamefont {Wan}},\ }\href@noop {} {\bibfield
  {journal} {\bibinfo  {journal} {Adv. Mater.}\ }\textbf {\bibinfo {volume}
  {35}},\ \bibinfo {pages} {2305609} (\bibinfo {year} {2023})}\BibitemShut
  {NoStop}%
\bibitem [{\citenamefont {Bhardwaj}, \citenamefont {Paasio},\ and\
  \citenamefont {Majumdar}(2025)}]{bhardwaj2025toward}%
  \BibitemOpen
  \bibfield  {author} {\bibinfo {author} {\bibfnamefont {K.}~\bibnamefont
  {Bhardwaj}}, \bibinfo {author} {\bibfnamefont {E.}~\bibnamefont {Paasio}}, \
  and\ \bibinfo {author} {\bibfnamefont {S.}~\bibnamefont {Majumdar}},\
  }\href@noop {} {\bibfield  {journal} {\bibinfo  {journal} {Advanced
  Intelligent Discovery}\ ,\ \bibinfo {pages} {e202500143}} (\bibinfo {year}
  {2025})}\BibitemShut {NoStop}%
\bibitem [{\citenamefont {Kim}\ \emph {et~al.}(2025)\citenamefont {Kim},
  \citenamefont {Kim}, \citenamefont {Kim}, \citenamefont {Kim}, \citenamefont
  {Lee}, \citenamefont {Chae}, \citenamefont {Je}, \citenamefont {Koo},\ and\
  \citenamefont {Kim}}]{kim2025voltage}%
  \BibitemOpen
  \bibfield  {author} {\bibinfo {author} {\bibfnamefont {J.~N.}\ \bibnamefont
  {Kim}}, \bibinfo {author} {\bibfnamefont {Y.~W.}\ \bibnamefont {Kim}},
  \bibinfo {author} {\bibfnamefont {B.}~\bibnamefont {Kim}}, \bibinfo {author}
  {\bibfnamefont {D.-H.}\ \bibnamefont {Kim}}, \bibinfo {author} {\bibfnamefont
  {G.~S.}\ \bibnamefont {Lee}}, \bibinfo {author} {\bibfnamefont {D.-H.}\
  \bibnamefont {Chae}}, \bibinfo {author} {\bibfnamefont {M.}~\bibnamefont
  {Je}}, \bibinfo {author} {\bibfnamefont {M.}~\bibnamefont {Koo}}, \ and\
  \bibinfo {author} {\bibfnamefont {Y.}~\bibnamefont {Kim}},\ }\href@noop {}
  {\bibfield  {journal} {\bibinfo  {journal} {Adv. Intell. Syst.}\ ,\ \bibinfo
  {pages} {2500028}} (\bibinfo {year} {2025})}\BibitemShut {NoStop}%
\bibitem [{\citenamefont {Yang}\ \emph {et~al.}(2016)\citenamefont {Yang},
  \citenamefont {Noh}, \citenamefont {Baek}, \citenamefont {Zheng},
  \citenamefont {Kang}, \citenamefont {Lee},\ and\ \citenamefont
  {Yoon}}]{yang2016memcapacitive}%
  \BibitemOpen
  \bibfield  {author} {\bibinfo {author} {\bibfnamefont {P.}~\bibnamefont
  {Yang}}, \bibinfo {author} {\bibfnamefont {Y.~J.}\ \bibnamefont {Noh}},
  \bibinfo {author} {\bibfnamefont {Y.-J.}\ \bibnamefont {Baek}}, \bibinfo
  {author} {\bibfnamefont {H.}~\bibnamefont {Zheng}}, \bibinfo {author}
  {\bibfnamefont {C.~J.}\ \bibnamefont {Kang}}, \bibinfo {author}
  {\bibfnamefont {H.~H.}\ \bibnamefont {Lee}}, \ and\ \bibinfo {author}
  {\bibfnamefont {T.-S.}\ \bibnamefont {Yoon}},\ }\href
  {http://dx.doi.org/10.1063/1.4941548} {\bibfield  {journal} {\bibinfo
  {journal} {Appl. Phys. Lett.}\ }\textbf {\bibinfo {volume} {108}} (\bibinfo
  {year} {2016})}\BibitemShut {NoStop}%
\bibitem [{\citenamefont {Park}\ \emph {et~al.}(2018)\citenamefont {Park},
  \citenamefont {Yang}, \citenamefont {Kim}, \citenamefont {Beom},
  \citenamefont {Lee}, \citenamefont {Kang},\ and\ \citenamefont
  {Yoon}}]{park2018analog}%
  \BibitemOpen
  \bibfield  {author} {\bibinfo {author} {\bibfnamefont {D.}~\bibnamefont
  {Park}}, \bibinfo {author} {\bibfnamefont {P.}~\bibnamefont {Yang}}, \bibinfo
  {author} {\bibfnamefont {H.~J.}\ \bibnamefont {Kim}}, \bibinfo {author}
  {\bibfnamefont {K.}~\bibnamefont {Beom}}, \bibinfo {author} {\bibfnamefont
  {H.~H.}\ \bibnamefont {Lee}}, \bibinfo {author} {\bibfnamefont {C.~J.}\
  \bibnamefont {Kang}}, \ and\ \bibinfo {author} {\bibfnamefont {T.-S.}\
  \bibnamefont {Yoon}},\ }\href {https://doi.org/10.1063/1.5043275} {\bibfield
  {journal} {\bibinfo  {journal} {Appl. Phys. Lett.}\ }\textbf {\bibinfo
  {volume} {113}} (\bibinfo {year} {2018})}\BibitemShut {NoStop}%
\bibitem [{\citenamefont {Emara}, \citenamefont {Aboudina},\ and\ \citenamefont
  {Fahmy}(2017)}]{emara2017non}%
  \BibitemOpen
  \bibfield  {author} {\bibinfo {author} {\bibfnamefont {A.~A.}\ \bibnamefont
  {Emara}}, \bibinfo {author} {\bibfnamefont {M.~M.}\ \bibnamefont {Aboudina}},
  \ and\ \bibinfo {author} {\bibfnamefont {H.~A.}\ \bibnamefont {Fahmy}},\
  }\href {\doibase https://doi.org/10.1016/j.mejo.2017.04.005} {\bibfield
  {journal} {\bibinfo  {journal} {Microelectron. J.}\ }\textbf {\bibinfo
  {volume} {64}},\ \bibinfo {pages} {39} (\bibinfo {year} {2017})}\BibitemShut
  {NoStop}%
\bibitem [{\citenamefont {Wang}\ \emph {et~al.}(2018)\citenamefont {Wang},
  \citenamefont {Rao}, \citenamefont {Han}, \citenamefont {Zhang},
  \citenamefont {Lin}, \citenamefont {Li}, \citenamefont {Li}, \citenamefont
  {Song}, \citenamefont {Asapu}, \citenamefont {Midya} \emph
  {et~al.}}]{wang2018capacitive}%
  \BibitemOpen
  \bibfield  {author} {\bibinfo {author} {\bibfnamefont {Z.}~\bibnamefont
  {Wang}}, \bibinfo {author} {\bibfnamefont {M.}~\bibnamefont {Rao}}, \bibinfo
  {author} {\bibfnamefont {J.-W.}\ \bibnamefont {Han}}, \bibinfo {author}
  {\bibfnamefont {J.}~\bibnamefont {Zhang}}, \bibinfo {author} {\bibfnamefont
  {P.}~\bibnamefont {Lin}}, \bibinfo {author} {\bibfnamefont {Y.}~\bibnamefont
  {Li}}, \bibinfo {author} {\bibfnamefont {C.}~\bibnamefont {Li}}, \bibinfo
  {author} {\bibfnamefont {W.}~\bibnamefont {Song}}, \bibinfo {author}
  {\bibfnamefont {S.}~\bibnamefont {Asapu}}, \bibinfo {author} {\bibfnamefont
  {R.}~\bibnamefont {Midya}},  \emph {et~al.},\ }\href {\doibase
  https://doi.org/10.1038/s41467-018-05677-5} {\bibfield  {journal} {\bibinfo
  {journal} {Nat. Commun.}\ }\textbf {\bibinfo {volume} {9}},\ \bibinfo {pages}
  {3208} (\bibinfo {year} {2018})}\BibitemShut {NoStop}%
\bibitem [{\citenamefont {Zheng}\ \emph {et~al.}(2019)\citenamefont {Zheng},
  \citenamefont {Wang}, \citenamefont {Gong}, \citenamefont {Yu}, \citenamefont
  {Chen}, \citenamefont {Cai}, \citenamefont {Huang}, \citenamefont {Jiang},
  \citenamefont {Xia},\ and\ \citenamefont {Huang}}]{8733844}%
  \BibitemOpen
  \bibfield  {author} {\bibinfo {author} {\bibfnamefont {Q.}~\bibnamefont
  {Zheng}}, \bibinfo {author} {\bibfnamefont {Z.}~\bibnamefont {Wang}},
  \bibinfo {author} {\bibfnamefont {N.}~\bibnamefont {Gong}}, \bibinfo {author}
  {\bibfnamefont {Z.}~\bibnamefont {Yu}}, \bibinfo {author} {\bibfnamefont
  {C.}~\bibnamefont {Chen}}, \bibinfo {author} {\bibfnamefont {Y.}~\bibnamefont
  {Cai}}, \bibinfo {author} {\bibfnamefont {Q.}~\bibnamefont {Huang}}, \bibinfo
  {author} {\bibfnamefont {H.}~\bibnamefont {Jiang}}, \bibinfo {author}
  {\bibfnamefont {Q.}~\bibnamefont {Xia}}, \ and\ \bibinfo {author}
  {\bibfnamefont {R.}~\bibnamefont {Huang}},\ }\href {\doibase
  10.1109/LED.2019.2921737} {\bibfield  {journal} {\bibinfo  {journal} {IEEE
  Electron Device Lett.}\ }\textbf {\bibinfo {volume} {40}},\ \bibinfo {pages}
  {1309} (\bibinfo {year} {2019})}\BibitemShut {NoStop}%
\bibitem [{\citenamefont {Lai}\ \emph {et~al.}(2010)\citenamefont {Lai},
  \citenamefont {Zhang}, \citenamefont {Li}, \citenamefont {Stickle},
  \citenamefont {Williams},\ and\ \citenamefont {Chen}}]{lai2010ionic}%
  \BibitemOpen
  \bibfield  {author} {\bibinfo {author} {\bibfnamefont {Q.}~\bibnamefont
  {Lai}}, \bibinfo {author} {\bibfnamefont {L.}~\bibnamefont {Zhang}}, \bibinfo
  {author} {\bibfnamefont {Z.}~\bibnamefont {Li}}, \bibinfo {author}
  {\bibfnamefont {W.~F.}\ \bibnamefont {Stickle}}, \bibinfo {author}
  {\bibfnamefont {R.~S.}\ \bibnamefont {Williams}}, \ and\ \bibinfo {author}
  {\bibfnamefont {Y.}~\bibnamefont {Chen}},\ }\href {\doibase
  https://doi.org/10.1002/adma.201000282} {\bibfield  {journal} {\bibinfo
  {journal} {Adv. Mater.}\ }\textbf {\bibinfo {volume} {22}},\ \bibinfo {pages}
  {2448} (\bibinfo {year} {2010})}\BibitemShut {NoStop}%
\bibitem [{\citenamefont {Hwang}\ \emph {et~al.}(2023)\citenamefont {Hwang},
  \citenamefont {Yu}, \citenamefont {Song}, \citenamefont {Hwang},\ and\
  \citenamefont {Kim}}]{hwang2023memcapacitor}%
  \BibitemOpen
  \bibfield  {author} {\bibinfo {author} {\bibfnamefont {S.}~\bibnamefont
  {Hwang}}, \bibinfo {author} {\bibfnamefont {J.}~\bibnamefont {Yu}}, \bibinfo
  {author} {\bibfnamefont {M.~S.}\ \bibnamefont {Song}}, \bibinfo {author}
  {\bibfnamefont {H.}~\bibnamefont {Hwang}}, \ and\ \bibinfo {author}
  {\bibfnamefont {H.}~\bibnamefont {Kim}},\ }\href {\doibase
  10.1002/advs.202303817} {\bibfield  {journal} {\bibinfo  {journal} {Adv.
  Sci.}\ }\textbf {\bibinfo {volume} {10}},\ \bibinfo {pages} {2303817}
  (\bibinfo {year} {2023})}\BibitemShut {NoStop}%
\bibitem [{\citenamefont {Kwon}\ and\ \citenamefont {Chung}(2020)}]{8970565}%
  \BibitemOpen
  \bibfield  {author} {\bibinfo {author} {\bibfnamefont {D.}~\bibnamefont
  {Kwon}}\ and\ \bibinfo {author} {\bibfnamefont {I.-Y.}\ \bibnamefont
  {Chung}},\ }\href {\doibase 10.1109/LED.2020.2969695} {\bibfield  {journal}
  {\bibinfo  {journal} {IEEE Electron Device Lett.}\ }\textbf {\bibinfo
  {volume} {41}},\ \bibinfo {pages} {493} (\bibinfo {year} {2020})}\BibitemShut
  {NoStop}%
\bibitem [{\citenamefont {Chen}\ and\ \citenamefont
  {Hwu}(2014)}]{chen2014effect}%
  \BibitemOpen
  \bibfield  {author} {\bibinfo {author} {\bibfnamefont {T.-Y.}\ \bibnamefont
  {Chen}}\ and\ \bibinfo {author} {\bibfnamefont {J.-G.}\ \bibnamefont {Hwu}},\
  }\href {https://link.springer.com/article/10.1007/s00339-014-8375-6}
  {\bibfield  {journal} {\bibinfo  {journal} {Appl. Phys. A}\ }\textbf
  {\bibinfo {volume} {116}},\ \bibinfo {pages} {1971} (\bibinfo {year}
  {2014})}\BibitemShut {NoStop}%
\bibitem [{\citenamefont {Li}, \citenamefont {Yang},\ and\ \citenamefont
  {Hwu}(2019)}]{8608000}%
  \BibitemOpen
  \bibfield  {author} {\bibinfo {author} {\bibfnamefont {H.-J.}\ \bibnamefont
  {Li}}, \bibinfo {author} {\bibfnamefont {C.-F.}\ \bibnamefont {Yang}}, \ and\
  \bibinfo {author} {\bibfnamefont {J.-G.}\ \bibnamefont {Hwu}},\ }\href
  {\doibase 10.1109/TED.2018.2889521} {\bibfield  {journal} {\bibinfo
  {journal} {IEEE Trans. Electron. Devices.}\ }\textbf {\bibinfo {volume}
  {66}},\ \bibinfo {pages} {1249} (\bibinfo {year} {2019})}\BibitemShut
  {NoStop}%
\bibitem [{\citenamefont {Kao}\ and\ \citenamefont
  {Hwu}(2025)}]{kao2025inversion}%
  \BibitemOpen
  \bibfield  {author} {\bibinfo {author} {\bibfnamefont {C.-Y.}\ \bibnamefont
  {Kao}}\ and\ \bibinfo {author} {\bibfnamefont {J.-G.}\ \bibnamefont {Hwu}},\
  }\href {https://doi.org/10.1063/5.0257074} {\bibfield  {journal} {\bibinfo
  {journal} {Appl. Phys. Lett.}\ }\textbf {\bibinfo {volume} {126}} (\bibinfo
  {year} {2025})}\BibitemShut {NoStop}%
\bibitem [{\citenamefont {Cai}\ \emph {et~al.}(2019)\citenamefont {Cai},
  \citenamefont {Li}, \citenamefont {Xu}, \citenamefont {Feng}, \citenamefont
  {Zhong}, \citenamefont {Xu}, \citenamefont {Gao},\ and\ \citenamefont
  {Wang}}]{cai2019organic}%
  \BibitemOpen
  \bibfield  {author} {\bibinfo {author} {\bibfnamefont {J.-W.}\ \bibnamefont
  {Cai}}, \bibinfo {author} {\bibfnamefont {L.-X.}\ \bibnamefont {Li}},
  \bibinfo {author} {\bibfnamefont {C.}~\bibnamefont {Xu}}, \bibinfo {author}
  {\bibfnamefont {Y.}~\bibnamefont {Feng}}, \bibinfo {author} {\bibfnamefont
  {Y.-N.}\ \bibnamefont {Zhong}}, \bibinfo {author} {\bibfnamefont {J.-L.}\
  \bibnamefont {Xu}}, \bibinfo {author} {\bibfnamefont {X.}~\bibnamefont
  {Gao}}, \ and\ \bibinfo {author} {\bibfnamefont {S.-D.}\ \bibnamefont
  {Wang}},\ }\href {https://doi.org/10.1063/1.5080115} {\bibfield  {journal}
  {\bibinfo  {journal} {Appl. Phys. Lett.}\ }\textbf {\bibinfo {volume} {114}}
  (\bibinfo {year} {2019})}\BibitemShut {NoStop}%
\bibitem [{\citenamefont {Li}\ \emph {et~al.}(2022)\citenamefont {Li},
  \citenamefont {Cai}, \citenamefont {Zhong}, \citenamefont {Gao},
  \citenamefont {Xu},\ and\ \citenamefont {Wang}}]{9845432}%
  \BibitemOpen
  \bibfield  {author} {\bibinfo {author} {\bibfnamefont {L.-X.}\ \bibnamefont
  {Li}}, \bibinfo {author} {\bibfnamefont {J.-W.}\ \bibnamefont {Cai}},
  \bibinfo {author} {\bibfnamefont {Y.-N.}\ \bibnamefont {Zhong}}, \bibinfo
  {author} {\bibfnamefont {X.}~\bibnamefont {Gao}}, \bibinfo {author}
  {\bibfnamefont {J.-L.}\ \bibnamefont {Xu}}, \ and\ \bibinfo {author}
  {\bibfnamefont {S.-D.}\ \bibnamefont {Wang}},\ }\href {\doibase
  10.1109/LED.2022.3195237} {\bibfield  {journal} {\bibinfo  {journal} {IEEE
  Electron Device Lett.}\ }\textbf {\bibinfo {volume} {43}},\ \bibinfo {pages}
  {1539} (\bibinfo {year} {2022})}\BibitemShut {NoStop}%
\bibitem [{\citenamefont {Ohtomo}\ and\ \citenamefont
  {Hwang}(2004)}]{ohtomo2004high}%
  \BibitemOpen
  \bibfield  {author} {\bibinfo {author} {\bibfnamefont {A.}~\bibnamefont
  {Ohtomo}}\ and\ \bibinfo {author} {\bibfnamefont {H.}~\bibnamefont {Hwang}},\
  }\href {\doibase https://doi.org/10.1038/nature02308} {\bibfield  {journal}
  {\bibinfo  {journal} {Nature}\ }\textbf {\bibinfo {volume} {427}},\ \bibinfo
  {pages} {423} (\bibinfo {year} {2004})}\BibitemShut {NoStop}%
\bibitem [{\citenamefont {Li}\ \emph {et~al.}(2011)\citenamefont {Li},
  \citenamefont {Richter}, \citenamefont {Paetel}, \citenamefont {Kopp},
  \citenamefont {Mannhart},\ and\ \citenamefont {Ashoori}}]{li2011very}%
  \BibitemOpen
  \bibfield  {author} {\bibinfo {author} {\bibfnamefont {L.}~\bibnamefont
  {Li}}, \bibinfo {author} {\bibfnamefont {C.}~\bibnamefont {Richter}},
  \bibinfo {author} {\bibfnamefont {S.}~\bibnamefont {Paetel}}, \bibinfo
  {author} {\bibfnamefont {T.}~\bibnamefont {Kopp}}, \bibinfo {author}
  {\bibfnamefont {J.}~\bibnamefont {Mannhart}}, \ and\ \bibinfo {author}
  {\bibfnamefont {R.}~\bibnamefont {Ashoori}},\ }\href {\doibase
  https://www.science.org/doi/10.1126/science.120416} {\bibfield  {journal}
  {\bibinfo  {journal} {Science}\ }\textbf {\bibinfo {volume} {332}},\ \bibinfo
  {pages} {825} (\bibinfo {year} {2011})}\BibitemShut {NoStop}%
\bibitem [{\citenamefont {Bi}\ \emph {et~al.}(2016)\citenamefont {Bi},
  \citenamefont {Huang}, \citenamefont {Bark}, \citenamefont {Ryu},
  \citenamefont {Lee}, \citenamefont {Eom}, \citenamefont {Irvin},\ and\
  \citenamefont {Levy}}]{bi2016electro}%
  \BibitemOpen
  \bibfield  {author} {\bibinfo {author} {\bibfnamefont {F.}~\bibnamefont
  {Bi}}, \bibinfo {author} {\bibfnamefont {M.}~\bibnamefont {Huang}}, \bibinfo
  {author} {\bibfnamefont {C.-W.}\ \bibnamefont {Bark}}, \bibinfo {author}
  {\bibfnamefont {S.}~\bibnamefont {Ryu}}, \bibinfo {author} {\bibfnamefont
  {S.}~\bibnamefont {Lee}}, \bibinfo {author} {\bibfnamefont {C.-B.}\
  \bibnamefont {Eom}}, \bibinfo {author} {\bibfnamefont {P.}~\bibnamefont
  {Irvin}}, \ and\ \bibinfo {author} {\bibfnamefont {J.}~\bibnamefont {Levy}},\
  }\href {https://doi.org/10.1063/1.4940045} {\bibfield  {journal} {\bibinfo
  {journal} {J. Appl. Phys.}\ }\textbf {\bibinfo {volume} {119}} (\bibinfo
  {year} {2016})}\BibitemShut {NoStop}%
\bibitem [{\citenamefont {Wu}\ \emph {et~al.}(2013)\citenamefont {Wu},
  \citenamefont {Wu}, \citenamefont {Qing}, \citenamefont {Zhou}, \citenamefont
  {Bao}, \citenamefont {Yang},\ and\ \citenamefont {Li}}]{wu2013electrically}%
  \BibitemOpen
  \bibfield  {author} {\bibinfo {author} {\bibfnamefont {S.}~\bibnamefont
  {Wu}}, \bibinfo {author} {\bibfnamefont {G.}~\bibnamefont {Wu}}, \bibinfo
  {author} {\bibfnamefont {J.}~\bibnamefont {Qing}}, \bibinfo {author}
  {\bibfnamefont {X.}~\bibnamefont {Zhou}}, \bibinfo {author} {\bibfnamefont
  {D.}~\bibnamefont {Bao}}, \bibinfo {author} {\bibfnamefont {G.}~\bibnamefont
  {Yang}}, \ and\ \bibinfo {author} {\bibfnamefont {S.}~\bibnamefont {Li}},\
  }\href {\doibase https://doi.org/10.1038/am.2013.48} {\bibfield  {journal}
  {\bibinfo  {journal} {NPG Asia Materials}\ }\textbf {\bibinfo {volume} {5}},\
  \bibinfo {pages} {e65} (\bibinfo {year} {2013})}\BibitemShut {NoStop}%
\bibitem [{\citenamefont {Kim}\ \emph {et~al.}(2015)\citenamefont {Kim},
  \citenamefont {Kim}, \citenamefont {Lim}, \citenamefont {Jeong},
  \citenamefont {Kwon}, \citenamefont {Baek},\ and\ \citenamefont
  {Kim}}]{kim2015electric}%
  \BibitemOpen
  \bibfield  {author} {\bibinfo {author} {\bibfnamefont {S.~K.}\ \bibnamefont
  {Kim}}, \bibinfo {author} {\bibfnamefont {S.-I.}\ \bibnamefont {Kim}},
  \bibinfo {author} {\bibfnamefont {H.}~\bibnamefont {Lim}}, \bibinfo {author}
  {\bibfnamefont {D.~S.}\ \bibnamefont {Jeong}}, \bibinfo {author}
  {\bibfnamefont {B.}~\bibnamefont {Kwon}}, \bibinfo {author} {\bibfnamefont
  {S.-H.}\ \bibnamefont {Baek}}, \ and\ \bibinfo {author} {\bibfnamefont
  {J.-S.}\ \bibnamefont {Kim}},\ }\href {\doibase
  https://doi.org/10.1038/srep08023} {\bibfield  {journal} {\bibinfo  {journal}
  {Sci. Rep.}\ }\textbf {\bibinfo {volume} {5}},\ \bibinfo {pages} {8023}
  (\bibinfo {year} {2015})}\BibitemShut {NoStop}%
\bibitem [{\citenamefont {Pradhan}\ \emph {et~al.}(2025)\citenamefont
  {Pradhan}, \citenamefont {Miller}, \citenamefont {Hartmann}, \citenamefont
  {Spring}, \citenamefont {Gabel}, \citenamefont {Leikert}, \citenamefont
  {Kuhn}, \citenamefont {Kamp}, \citenamefont {Lopez-Richard}, \citenamefont
  {Sing} \emph {et~al.}}]{pradhan2025oxide}%
  \BibitemOpen
  \bibfield  {author} {\bibinfo {author} {\bibfnamefont {S.}~\bibnamefont
  {Pradhan}}, \bibinfo {author} {\bibfnamefont {K.}~\bibnamefont {Miller}},
  \bibinfo {author} {\bibfnamefont {F.}~\bibnamefont {Hartmann}}, \bibinfo
  {author} {\bibfnamefont {M.}~\bibnamefont {Spring}}, \bibinfo {author}
  {\bibfnamefont {J.}~\bibnamefont {Gabel}}, \bibinfo {author} {\bibfnamefont
  {B.}~\bibnamefont {Leikert}}, \bibinfo {author} {\bibfnamefont
  {S.}~\bibnamefont {Kuhn}}, \bibinfo {author} {\bibfnamefont {M.}~\bibnamefont
  {Kamp}}, \bibinfo {author} {\bibfnamefont {V.}~\bibnamefont {Lopez-Richard}},
  \bibinfo {author} {\bibfnamefont {M.}~\bibnamefont {Sing}},  \emph {et~al.},\
  }\href@noop {} {\bibfield  {journal} {\bibinfo  {journal} {arXiv preprint
  arXiv:2508.03515}\ } (\bibinfo {year} {2025})}\BibitemShut {NoStop}%
\bibitem [{\citenamefont {Nicollian}\ and\ \citenamefont
  {Goetzberger}(1967)}]{Nicollian1967}%
  \BibitemOpen
  \bibfield  {author} {\bibinfo {author} {\bibfnamefont {E.~H.}\ \bibnamefont
  {Nicollian}}\ and\ \bibinfo {author} {\bibfnamefont {A.}~\bibnamefont
  {Goetzberger}},\ }\href {\doibase 10.1002/j.1538-7305.1967.tb01727.x}
  {\bibfield  {journal} {\bibinfo  {journal} {The Bell System Technical
  Journal}\ }\textbf {\bibinfo {volume} {46}},\ \bibinfo {pages} {1055}
  (\bibinfo {year} {1967})}\BibitemShut {NoStop}%
\bibitem [{\citenamefont {Lopez-Richard}\ \emph {et~al.}(2023)\citenamefont
  {Lopez-Richard}, \citenamefont {Silva}, \citenamefont {Lipan},\ and\
  \citenamefont {Hartmann}}]{LopezRichard2022}%
  \BibitemOpen
  \bibfield  {author} {\bibinfo {author} {\bibfnamefont {V.}~\bibnamefont
  {Lopez-Richard}}, \bibinfo {author} {\bibfnamefont {R.~S.~W.}\ \bibnamefont
  {Silva}}, \bibinfo {author} {\bibfnamefont {O.}~\bibnamefont {Lipan}}, \ and\
  \bibinfo {author} {\bibfnamefont {F.}~\bibnamefont {Hartmann}},\ }\href
  {\doibase 10.1063/5.0142721} {\bibfield  {journal} {\bibinfo  {journal} {J.
  Appl. Phys.}\ }\textbf {\bibinfo {volume} {133}},\ \bibinfo {pages} {134901}
  (\bibinfo {year} {2023})}\BibitemShut {NoStop}%
\end{thebibliography}%

\clearpage
\onecolumngrid

\begin{figure*}[ht]
\centering
\includegraphics[page=1,width=\textwidth]{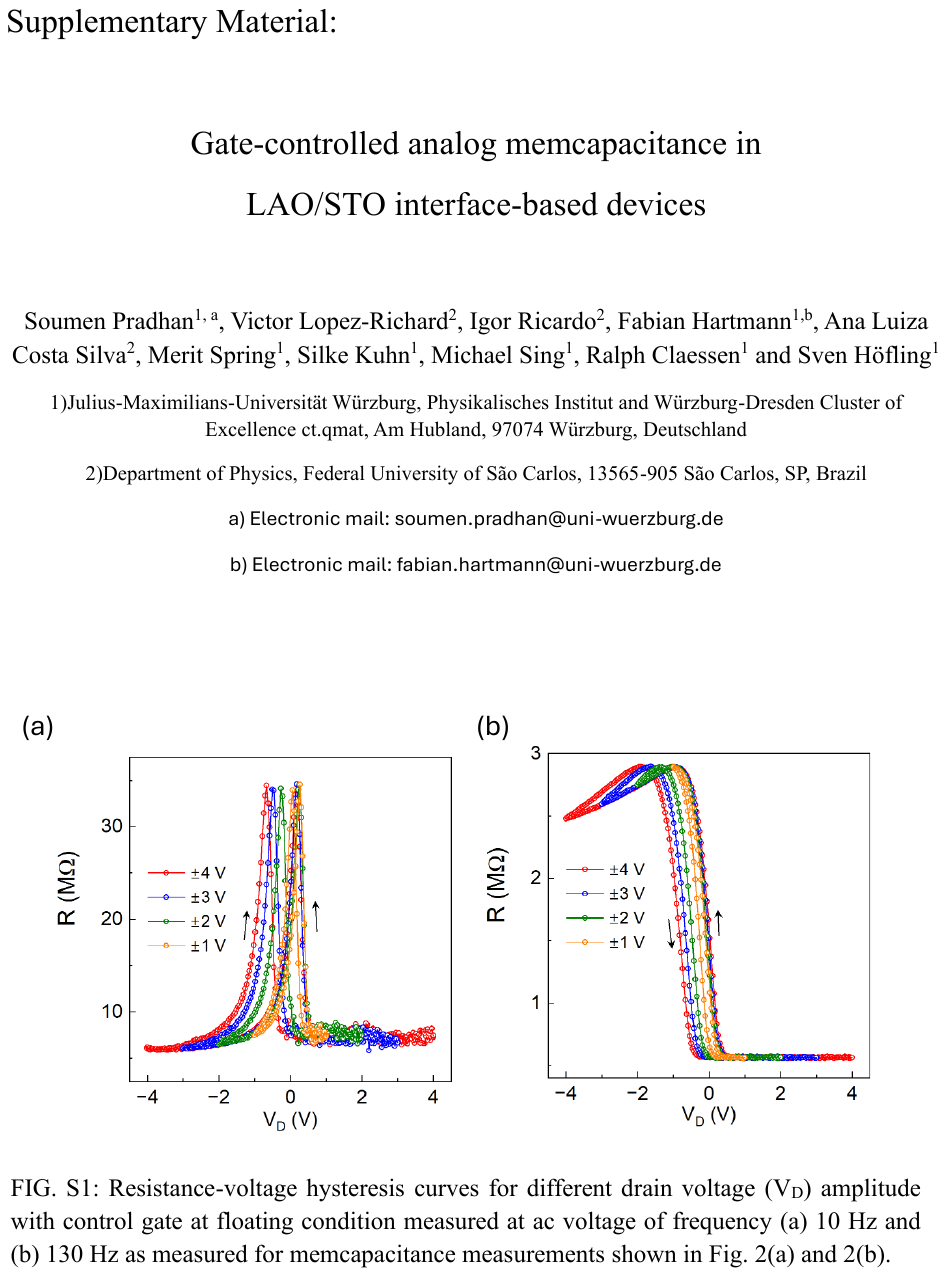}
\end{figure*}

\begin{figure*}[ht]
\centering
\includegraphics[page=2,width=\textwidth]{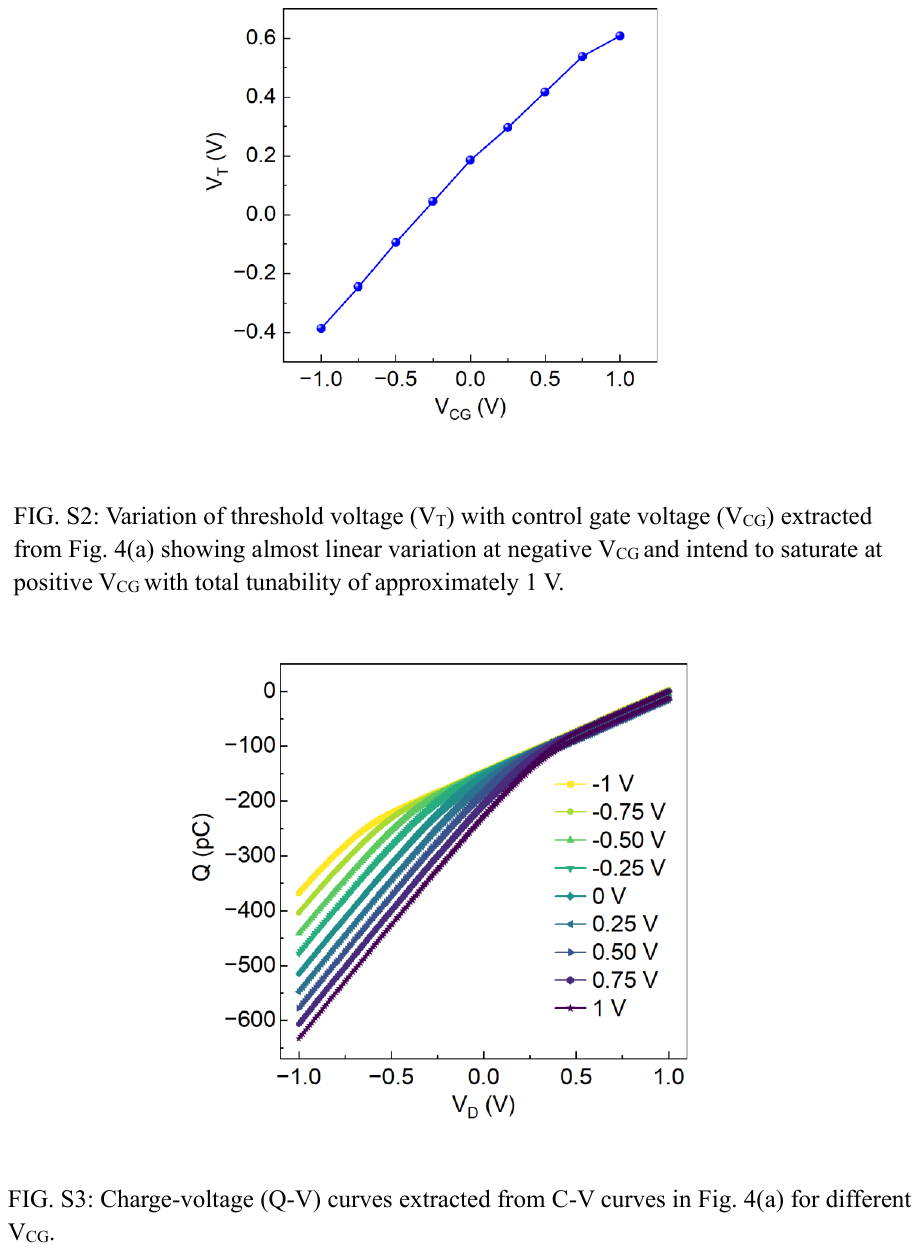}
\end{figure*}

\begin{figure*}[ht]
\centering
\includegraphics[page=3,width=\textwidth]{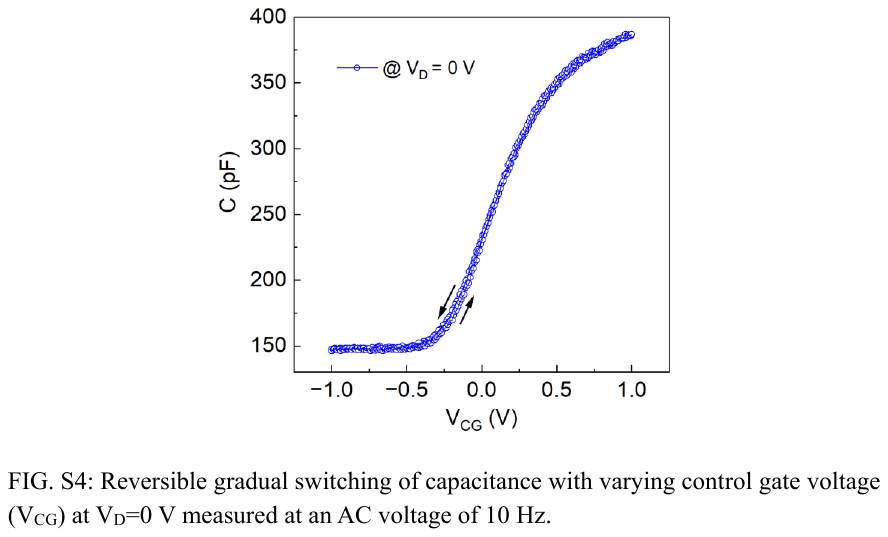}
\end{figure*}
\clearpage
\twocolumngrid
\end{document}